\documentclass[aps,preprintnumbers,showpacs,twocolumn,superscriptaddress,floatfix,nofootinbib]{revtex4-1}

\usepackage{latexsym} 
\usepackage{amssymb} 
\usepackage{amsmath} 
\usepackage{amsfonts}
\usepackage{bm}
\usepackage{color}
\usepackage{times} 
\usepackage{units}
\usepackage{hyperref}

\usepackage[utf8x]{inputenc}
\usepackage{graphicx} 
\usepackage[squaren]{SIunits} 
\usepackage[inline]{enumitem}

\graphicspath{ {./plots/} }

\newcommand{\Eref}[1]{Eq.~\eqref{#1}} 
\newcommand{\Fref}[1]{Fig.~\ref{#1}} 
\newcommand{\Sref}[1]{Sec.~\ref{#1}}

\begin{document}


\title{Finite tidal effects in GW170817: Observational evidence or model assumptions?}

\author{W. Kastaun}
\author{F. Ohme}

\address{Max Planck Institute for Gravitational Physics
(Albert Einstein Institute), Callinstr. 38, D-30167 Hannover, Germany}
\address{Leibniz Universität Hannover, D-30167 Hannover, Germany}


\begin{abstract}
After the detection of gravitational waves caused by the coalescence of compact 
objects in the mass range of neutron stars, GW170817, several studies have searched 
for an imprint of tidal effects in the signal, employing different model assumptions. 
One important distinction is whether or not to  
assume that both objects are neutron stars and obey the same equation of state.
Some studies assumed independent properties. Others assume a universal equation
of state, and in addition that the tidal deformability follows certain 
phenomenological relations. An important question is 
whether the gravitational-wave data alone constitute observational evidence for finite tidal effects.
All studies provide Bayesian credible intervals,
often without sufficiently discussing the impact of prior assumptions, especially in the case of lower limits on the neutron-star tidal deformability or radius.
In this article, we scrutinize the implicit and explicit prior assumptions made 
in those studies. Our findings strongly indicate that existing lower credible
bounds are mainly a consequence of prior assumptions combined with information gained about the system's masses. Importantly, those lower bounds are typically not informed by the observation of tidal effects in the gravitational-wave signal. Thus, regarding them as 
observational evidence might be misleading without a more detailed discussion. 
Further, we point out technical problems and conceptual 
inconsistencies in existing studies.
We also assess the limitations due to systematic waveform model uncertainties
in a novel way, demonstrating that differences between existing models are not guaranteed to be small enough for an unbiased estimation of lower bounds on the tidal deformability.
Finally, we propose strategies for gravitational-wave data analysis designed
to avoid some of the problems we uncovered. 
\end{abstract}


\maketitle

\section{Introduction}
\label{sec:intro}
Among known astronomical objects, the highest matter densities occur
within neutron stars (NSs).
The behavior of cold matter at supranuclear 
densities cannot be studied in laboratories on Earth. Therefore, 
fundamental properties such as the equation of state (EOS) that relates 
pressure, temperature, and density, are essentially unconstrained by experiments. 
Theoretical models calibrated by properties of atomic nuclei 
can be used to extrapolate the EOS of neutron-star matter, but face 
considerable theoretical uncertainties.
Astrophysical observations of NSs can be used instead to infer
EOS constraints, based on the EOS impact on NS structure and dynamics. 

Merging NS binaries in particular are an excellent environment to probe the EOS, 
since the gravitational wave (GW) signal emitted before, during, and after merger
carries signatures of the EOS, as do electromagnetic (EM) counterparts
given by kilonova and short gamma ray burst.
The first multimessenger observation  
\cite{LVC:BNSDetection, LVC:GWGRB:2017, LVC:MMA:2017} 
of a binary neutron star (BNS) merger event, GW170817,
indeed opened a wealth of opportunities.

The most direct observable signature of the EOS is contained in the GW signal
emitted during the last seconds before the merger.
While a neutron star and another compact object spiral around each other,
the neutron star is deformed
in the tidal field of the companion.
This deformation modifies the dissipation of energy and angular momentum by
GW and thereby affects the decay of the orbit.
Measuring the inspiral frequency evolution therefore carries imprints of the 
EOS \cite{Flanagan:2008:021502}, in addition to the constituents' masses and 
angular momenta.

For binary neutron stars, the frequencies during this phase fall within the 
sensitive range of current ground-based interferometers, such as the 
Advanced Laser Interferometer Gravitational-Wave Observatory (LIGO)  \cite{AdvLIGO:2015}
or the Virgo detector  \cite{Virgo2015}.
In contrast, merger and postmerger GW emission is dominated by 
frequencies $>1\,\textrm{kHz}$, for which the current detection range 
is considerably smaller, making observations of this phase unlikely (compare 
\cite{LVC:BNSSourceProp:2019,LVC:PM:2017,LVC:PMLong:2019}).

For the first observation of GWs from a binary NS merger, GW170817 \cite{LVC:BNSDetection}, 
several studies have already presented constraints on the tidal parameters that enter the GW 
signal  \cite{LVC:BNSSourceProp:2019, LVC:EOSPaper:2018, SoumiDe:2018, Landry:2018prl, Capano:2019eae:arxiv}. 
All of those provide robust upper limits on the tidal effects.
Although more difficult to measure, lower bounds are provided as well.

In this paper, we revisit the methods used in  
\cite{LVC:BNSSourceProp:2019, LVC:EOSPaper:2018, SoumiDe:2018} to derive such constraints.
We scrutinize the implicit and explicit assumptions involved, in order to decide
if the published lower limits can be interpreted as direct observational 
evidence of finite tidal effects, or rather constitute inferences based on
model assumptions and observational constraints on the masses.
The latter would be the case when the same (or very similar) limits are obtained regardless 
of whether the real source was a neutron star or black hole (BH) binary.

We emphasize that we do not argue against combining observational data and prior knowledge,
e.g., from nuclear physics, in order to provide tighter constraints on the EOS.
Before interpreting such constraints as observational evidence, however, 
the contribution of model assumptions needs to be accounted for. In any case,
if the model assumptions include ambiguous \textit{ad-hoc} choices or inconsistencies,
then resulting constraints are only meaningful to the extent that they are dominated by the 
observational data and not the prior knowledge. 
We review prior choices with regard to the above.

Before explaining the main challenges, we briefly review the 
Bayesian framework that is typically used to infer the properties of binary GW 
sources (see \cite{Veitch:2009hd}). The binary's vector of parameters, $\theta$  (composed of the binary NS 
properties such as masses, distance, inclination, 
etc., and also EOS-related parameters) is determined in the form of its posterior 
probability density, 
\begin{equation}
p(\theta \vert d, I) = \frac{P(d \vert \theta, I)}{P(d \vert I)} \; p(\theta 
\vert I), \label{eq:Bayes_law}
\end{equation}
given the observed data $d$ and any 
additional model assumptions $I$. Bayes' Law, \Eref{eq:Bayes_law}, relates the 
probability density distribution on the left-hand side to the product 
of the likelihood of the observed the data given the source parameters, 
$P(d \vert \theta, I)$, and the prior probability density of encountering a system with 
those parameters, $p(\theta \vert I)$. The likelihood is calculated by comparing 
gravitational waveform models with the data, assuming stationary Gaussian 
noise. The prior represents any knowledge or belief that existed before the 
data were obtained.

The main conceptual challenges all such Bayesian EOS analyses of binary 
NS GW data face are the following:
\begin{enumerate}[itemsep=0pt]
 \item \emph{Parameterization of the unknown EOS.} GW models depend on two 
parameters related to the tidal deformability of each object in the binary. 
If a common but unknown EOS is assumed, relating those two numbers is nontrivial. 
In Secs.~\ref{sec:univ} and \ref{sec:real}, 
we discuss model assumptions  \cite{Yagi:2016:13LT01} (often 
referred to as ``universal relations'') that have been used to correlate 
masses and tidal deformabilities of NS binaries, analyze their range of 
applicability as well as pathologies arising from their use, and suggest a simpler
alternative.
\item \emph{Using the posterior to claim observational evidence for tidal effects.}
Even for a completely unconstraining measurement, Bayesian methods provide
a posterior distribution, which then only reflects the chosen prior distribution.
To assess the contribution of observational data, 
one has to compare posterior and prior. 
This becomes nontrivial in a multidimensional parameter space.
If the marginalized posterior for tidal effects is constrained in comparison
to the prior, it still does not signify a direct measurement of tidal effects. 
The reason is that the data might only constrain the NS masses. If the prior 
distribution for tidal effects depends on the masses, the posterior is then influenced
even without any constraint of tidal effects. We therefore separate
the indirect impact of observational constraints on quantities that are unrelated to 
tidal effects from the impact of those that are directly related.
For this, we discuss idealized cases such as fixed mass ratios.
\item \emph{Ambiguity of the prior probability for the EOS-related quantities.} 
The choice of prior is a well-known issue in Bayesian analysis. 
There are no other observations
that could be used as prior knowledge on tidal deformability, and the 
requirement to specify a prior distribution can only be satisfied by ambiguous 
\textit{ad-hoc} choices.
We examine the shape of priors used for published EOS constraints
in detail and discuss implications.
\item \emph{Credible bounds do not imply statements on likelihood of particular models.}
Depending on the shape of the posterior,
a specific model with parameters outside the credible bounds can be more likely
than another model within the bounds. Even the maximum likelihood parameter
can be located outside a given credible interval. It is always possible to 
compute a two-sided credible interval, which, on its own, is therefore not 
useful as evidence against specific models. 
For this, Bayesian model selection is more suitable. A recent model selection 
study on GW170817 \cite{LVC:EOSModelSel:2019} did not find evidence for or against 
finite tidal effects. In this work, we do not compute Bayes factors, but provide 
arguments, based on features of the posterior and prior distributions, 
against attributing the lower credible bounds to observational evidence.
\item \emph{Theoretical uncertainty of waveform model.}
Computing the likelihood requires a theoretical model of the gravitational 
waveform for given parameters. The systematic error of the latter is currently
not known, but estimated by comparing different theoretical models 
\cite{Favata:2014:101101}.
Further, estimating the resulting systematic error of credible bounds based on 
statistical analysis of a given event is complicated by the presence of 
the waveform-independent statistical errors.
In \Sref{sec:syserr} we provide a new perspective on estimating the potential impact 
of systematic waveform errors. The order of magnitude we find 
for the systematic errors 
indicates that lower bounds on tidal parameters should 
be interpreted with great caution.
\end{enumerate}

The issues discussed in this work are also relevant when planning the data 
analysis strategy for future events.
This involves a choice of how many model assumptions to incorporate already during the GW 
data analysis, and which quantities to provide for comparison to nuclear physics models.
In Sec.~\ref{sec:improved_analysis}, we propose a strategy that keeps the division 
line between data analysis and neutron-star modeling as close to the measurable 
quantities as possible, which simplifies the interpretation of results and
allows one to study the consequences of different model assumptions without repeating 
computationally expensive parameter estimation studies.
Employing a reasonable strategy will be important to facilitate 
the exciting synergy of GW astronomy and nuclear physics as more observations 
become available.

\section{Binary neutron-star parameters}
\label{sec:basics}

\subsection{Masses}

The evolution of the GW signal during the inspiral depends most strongly 
on a combination of the individual gravitational masses
$M_1$ and $M_2$ called chirp mass,
\begin{align}
M_c &= \frac{\left(M_1 M_2 \right)^{3/5}}{\left( M_1 + M_2 \right)^{1/5}}.
\end{align}
For a BNS merger, the chirp mass can be determined very accurately by current GW 
detectors. We therefore treat it as fixed in this article. We 
specifically use the value $M_c=1.186$ inferred for GW170817 \cite{LVC:BNSSourceProp:2019}.  
The mass ratio $q = M_2/M_1$ is less well constrained. 
For the given chirp mass, we can express the individual NS masses as a function 
of the mass ratio,
\begin{align}
M_1(q) &= M_c \left( 1 + q \right)^{1/5} q^{-3/5}, \\
M_2(q) &= M_c \left( 1 + q \right)^{1/5} q^{2/5}.
\end{align}
We focus on the simple case of vanishing NS spins in this work, 
although they can have a significant influence on the GW signal.
The particular challenges that we discuss are not reduced when allowing 
for spins, while ignoring spins considerably simplifies the discussion.
We also make the common assumption of negligible orbital eccentricity.

\subsection{Effective tidal deformability}
\label{sec:efftidal}

In the context of GW signal, the difference between NS and BH cases is 
how the objects' quadrupole moments react to the tidal forces caused by their 
companion. This is expressed by the dimensionless tidal deformabilies
$\Lambda_1$, $\Lambda_2$ of the two objects, which are 0 for BHs (see \cite{Gralla_2018} however).
Since we neglect the NS spins, finite size 
effects are given by the tidal deformability as a function of 
gravitational mass alone, $\Lambda(M)$, which depends on the EOS.
The correction to the point-particle GW signal mainly depends on 
a combination (see \cite{Flanagan:2008:021502}) called effective tidal deformability,
\begin{align}
\tilde{\Lambda} 
&= \frac{16}{13} 
   \frac{\left(1 + 12 q \right) \Lambda_1 + \left(q + 12 \right) q^4 \Lambda_2}
        {\left(1 + q \right)^5}. \label{eq:lambda_eff}
\end{align}
In this work, we use the rescaled quantity
\begin{align}\label{eq:lambda_bar}
\bar{\Lambda} &= \tilde{\Lambda} w_f^{-1}(q),  \\
w_f(q) &= \frac{16}{13} \frac{1 + 12 q + \left(q + 12 \right) q^4}{\left(1 + q \right)^5}. 
\end{align}
The interpretation of $\bar{\Lambda}$ is convenient because
for a BNS, it results from a linear interpolation of $\Lambda(M)$ between
the masses $M_1$ and $M_2$, evaluated at a mass $\bar{M}$ defined by
\begin{align}
\bar{M}(q) &= w_1(q) M_1(q) + w_2(q) M_2(q), \label{eq:m_interpol}\\
\bar{\Lambda}(q) &=  w_1(q) \Lambda(M_1(q)) + w_2(q) \Lambda(M_2(q)). \label{eq:l_interpol}  
\end{align}
For mass ratios close to unity, $\bar{\Lambda} \approx \Lambda(\bar{M})$.
The weights depend only on the mass ratio, as
\begin{align}
w_1(q) &= \frac{1 + 12 q}{1 + 12 q + \left(q + 12 \right) q^4}, \\
w_2(q) &= \frac{\left(q + 12 \right) q^4}{1 + 12 q + \left(q + 12 \right) q^4}.
\end{align}
where $w_1 + w_2 = 1$. We follow the convention $q<1$, and therefore 
$w_1\ge 1/2$ and $w_2 \le 1/2$.
Note $\bar{M}(q)$ is monotonically decreasing, which allows us to express $\bar{\Lambda}$ as 
a function of $\bar{M}$.

\begin{figure}
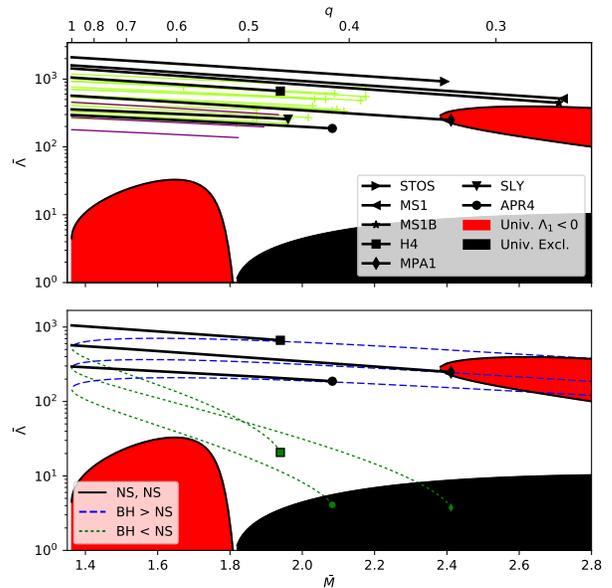

  \begin{center}
    \includegraphics[width=0.95\columnwidth]{{{Lambda_m_vs_M_m_all}}}  
    \caption{Top: 
    Interpolated tidal deformability $\bar{\Lambda}$ of a BNS system 
    as a function of $\bar{M}$ (bottom axis) and $q$ (top axis).
    The curves correspond to various EOS, where only a representative set is labeled
    and the others are shown in green. Causality-violating EOS used in \cite{LVC:EOSPaper:2018}
    are shown in purple.     
    The sequences terminate when $M_1(q)$ reaches the maximum mass (marked by symbols).
    The shaded areas are regions unreachable when using the universal relations from
    \cite{LVC:EOSPaper:2018}, either implying negative tidal deformabilites (red)
    or by construction (black).
    Bottom: Comparison of $\bar{\Lambda}(\bar{M})$ between the BNS case and mixed NS-BH case, 
    for three EOS.
    The solid curves refer to the BNS case, the dashed (dotted) curves to a 
    mixed NS/BH binary where the heavier (lighter) object is the BH. 
    }
    \label{fig:lambda_m_vs_M_m_all}
  \end{center}
\end{figure}

For the BNS case with both NS obeying the same EOS, $\bar{\Lambda}$ is shown in the upper panel of 
\Fref{fig:lambda_m_vs_M_m_all} for a representative set of EOS. Note that the function is defined 
only up to a maximum $\bar{M}$ (minimum $q$) for which $M_1$ exceeds the maximum 
allowed mass for a nonrotating NS for the given EOS. For the case of a mixed binary, we can 
compute $\bar{\Lambda}$ by setting either $\Lambda_1$ or $\Lambda_2$ to 0 in \Eref{eq:lambda_eff}.
The resulting differences are shown in the lower panel for \Fref{fig:lambda_m_vs_M_m_all}.

Although only $\bar{\Lambda}$ can be confined by the GW signal, the difference of the 
tidal deformabilities of the two NSs is an important property of the EOS as well.
In the following, we express it in terms of a dimensionless parameter
\begin{align}
\bar{S} &= \frac{\ln(\Lambda_1) - \ln(\Lambda_2)}{\ln(M_1) - \ln(M_2)}
        = \frac{\ln(\Lambda_2/\Lambda_1)}{\ln(q)}. \label{eq:def_sl}
\end{align}
For moderate mass ratios, $\bar{S}$ has an intuitive interpretation 
in terms of the slope of $\Lambda(M)$,

\begin{align}
\bar{S} \approx \frac{\mathrm{d} \ln(\Lambda)}{\mathrm{d}\ln(M)}(\bar{M}).
\end{align}

In the context of waveform models, it is more common to use a quantity 
$\delta\tilde{\Lambda}$, which enters at higher post-Newtonian order than 
$\tilde{\Lambda}$ and has minor impact on the waveform \cite{Favata:2014:101101}.
For fixed $\bar{\Lambda}$ and $q$, the quantity $\delta\tilde{\Lambda}$ 
becomes a function of $\bar{S}$, which therefore has minor impact as well. 
Unless noted otherwise, we discuss the idealized case
that the GW signal depends on
$M_c, q, \bar{\Lambda}$ but not $\bar{S}$.

\subsection{Universal relations}
\label{sec:univ}
In order to make statements about the individual tidal deformabilites,
the measurement of $\bar{\Lambda}$ is not sufficient; the slope $\bar{S}$ is 
required as well. Since it does not enter the GW signal, it cannot be 
measured. One way to infer properties of the individual NSs regardless of this fact
is to use model-dependent assumptions about $\bar{S}$.

As proposed in \cite{Yagi:2016:13LT01}, one can express the effective tidal 
deformability $\tilde{\Lambda}$ as an EOS-independent function of the average 
deformability 
and the mass ratio. 
The residual error (computed in \cite{Yagi:2016:13LT01})
of the expression in comparison 
to the results for a few nuclear physics EOS models
is small enough to motivate the term universal relation.

In detail, the relations are written as $\Lambda_a(\Lambda_s, q)$, where
$\Lambda_s = \left( \Lambda_1 + \Lambda_2 \right) / 2$ and
$\Lambda_a = \left( \Lambda_2 - \Lambda_1 \right) / 2$.
The function $\Lambda_a$ is expanded in $\Lambda_s$ and $q$ as 
\begin{align}\label{eq:univ}
\Lambda_a &= 
  \Lambda_s F_n(q) 
  \frac{1 + \sum_{l=1}^3 \sum_{k=1}^2 b_{lk} q^k \Lambda_s^{-l/5}}{
        1 + \sum_{l=1}^3 \sum_{k=1}^2 c_{lk} q^k \Lambda_s^{-l/5}}, \\
F_n(q) &= 
  \frac{1 - q^{10/(3-n)} }{ 1 + q^{10/(3-n)}} .
\end{align}
Those relations are used in \cite{LVC:EOSPaper:2018} for data analysis of GW170817. 
The coefficients $n, b_{ij}, c_{ij}$, given in \cite{Chatziioannou:2018:104036},
were obtained by fitting to a set of nuclear physics EOS. 
We noticed that the coefficients differ from those given in 
\cite{Yagi:2016:13LT01} (after taking into account that the 
definitions differ by inclusion of a scaling factor).
However, the qualitative statements we make on universal 
relation usage are equally valid for both.

The set of EOS used for calibration in 
\cite{LVC:EOSPaper:2018,Chatziioannou:2018:104036} consists 
of SLY \cite{2001AA...380..151D}, LS220\cite{LATTIMER:1991:331}, 
Shen \cite{SHEN:1998:435} (here:STOS), AP3 and AP4 \cite{PhysRevC.58.1804}, 
WFF1 and WFF2 \cite{1988PhRvC..38.1010W}, ENG \cite{1996ApJ...469..794E}, 
MPA1 \cite{1987PhLB..199..469M}, MS1 and MS1b \cite{1996NuPhA.606..508M}.
We note that the AP3, AP4, WFF1, WFF2, and ENG EOS start to violate causality before 
reaching the central density of the maximum mass nonrotating NS solution. 
It is also worth pointing out that the set of EOS models used for calibration has 
maximum NS masses compatible with the observation of PSR J0348+0432 
\cite{Antoniadis:2013:448} (although
for the WFF1 EOS, causality is violated at this mass). Since low tidal deformabilities
typically occur near the maximum mass model and increase with decreasing mass,
nuclear physics EOS candidates compatible with PSR J0348+0432 do not yield low 
deformabilities in the mass range measured for GW170817 (for the EOS set above,
$\Lambda>61$ in the mass range inferred for the low spin prior 
in \cite{LVC:BNSSourceProp:2019}; compare Fig.~1 in \cite{LVC:EOSModelSel:2019}).
On the other hand, strong phase transitions are not considered and might alter the 
behavior of tidal deformability.

\begin{figure}
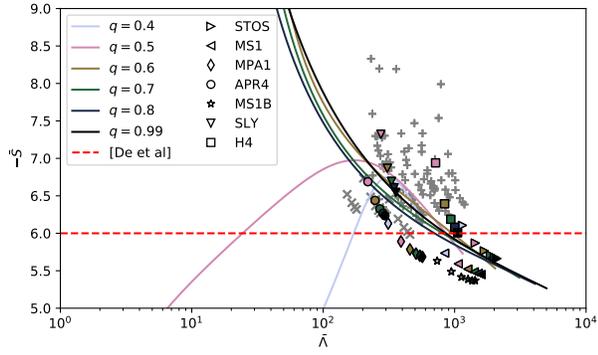

  \begin{center}
    \includegraphics[width=0.95\columnwidth]{{{Sl_vs_Lambda_m}}}  
    \caption{Universal relations from \cite{LVC:EOSPaper:2018, Chatziioannou:2018:104036}, 
    evaluated for fixed chirp mass. The solid curves show
    the parameters $-\bar{S}$, $\bar{\Lambda}$ resulting for fixed mass ratios, 
    varying $\Lambda_s$ up to $5000$ (the upper limit used in \cite{Chatziioannou:2018:104036}). 
    Note at very small $\Lambda_s$, there 
    is an additional branch with $\bar{S}>0$ (not shown) due to the poles 
    of the universal relations. 
    The dashed line represents $-\bar{S}$ from \cite{SoumiDe:2018}, which does not depend
    on $q$ or $\bar{\Lambda}$.
    The labeled symbols mark the values computed for a representative set of nuclear 
    physics EOS, colored according to mass ratio like the universal relation curves.
    Grey plus symbols show additional EOS, 
    and grey crosses mark causality-violating EOS used in 
    \cite{LVC:EOSPaper:2018,Chatziioannou:2018:104036}.}
    \label{fig:Sl_vs_Lambda_m}
  \end{center}
\end{figure}

The universal relations $\Lambda_a(\Lambda_s, q)$ are shown in \Fref{fig:Sl_vs_Lambda_m},
in terms of our $\bar{\Lambda}$, $\bar{S}$ parametrization.
The relations are evaluated for our fixed fiducial chirp mass, and various mass ratios in 
the range $0.4\ldots 1$.
For comparison, we plot the values $\bar{\Lambda}, \bar{S}$ for a representative set of EOS. 
As one can see, none of the EOS reach low values of $\bar{\Lambda}$ for the mass range of the 
detected event. Using the universal relations at lower values of $\bar{\Lambda}$
is self-contradictory, since the EOS used to derive the relations explicitly exclude
such values. We note a subtle point: the models used in 
\cite{Yagi:2016:13LT01} to fit the universal relations do include lower values 
of $\bar{\Lambda}$. The reason is that the mass range of the models for the fit 
is not restricted. One should however keep in mind that the resulting fit around 
any given $\bar{\Lambda}$ is only valid within a range for $\bar{M}$ allowed by the 
EOS models. In other words, there might be EOS leading to low $\bar{\Lambda}$ for the 
detected mass range, but these EOS must then be very different from those used in the fit. 
In turn, the error of $\bar{S}$ from the universal relations is completely unknown in that 
case.

\begin{figure}
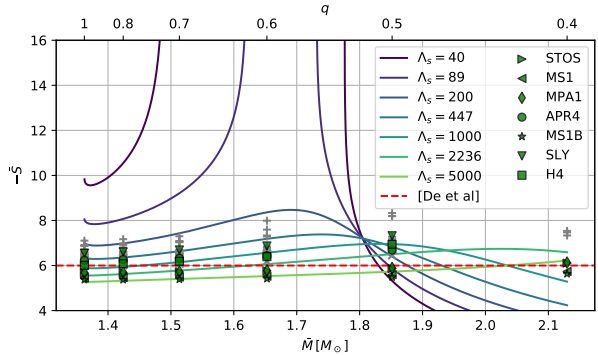

  \begin{center}
    \includegraphics[width=0.95\columnwidth]{{{Sl_vs_M_m}}}  
    \caption{Universal relations from 
    \cite{LVC:EOSPaper:2018,Chatziioannou:2018:104036}, evaluated for fixed 
    chirp mass. The curves show the slope parameter $\bar{S}$ versus 
    $\bar{M}$ for fixed values of $\Lambda_s$ and varying 
    mass ratio. The curves diverge where $\Lambda_1$ becomes negative.
    The symbols mark the values obtained for 
    the same EOS as in \Fref{fig:Sl_vs_Lambda_m}.
    }
    \label{fig:Sl_vs_M_m}
  \end{center}
\end{figure}

Besides the conceptual problem above, the relations used in \cite{LVC:EOSPaper:2018}
exhibit some technical problems.
For low values of $\Lambda_s$, the expression \Eref{eq:univ} has 
a pole and ceases to be meaningful. The corresponding slope $\bar{S}$ diverges.
The divergence leads to gaps in the possible 
values of $\bar{\Lambda}$. However, the divergence only affects very small values.
More important is the fact that $\Lambda_1$ computed from the universal relation 
crosses 0 (we recall that negative tidal deformabilities are unphysical 
\cite{Tianqi:2018:063020}). At this point $\bar{S}$ diverges since it is defined in a way
that does not allow negative deformabilities.
This happens below some critical $\bar{\Lambda}$ that depends on the mass 
ratio. Even before $\Lambda_1$ crosses 0, the slope $\bar{S}$ greatly exceeds the
range given by the EOS models, as shown in \Fref{fig:Sl_vs_M_m}.
Assuming the universal relations to hold universally thus excludes
small effective tidal deformabilities. The excluded region is shown in 
\Fref{fig:lambda_m_vs_M_m_all} as a function of $\bar{M}$.
Although not relevant for the case of GW170817, we note that for mass ratios
$q\lesssim 0.5$, low values of $\bar{\Lambda}$ are excluded because the universal
relation develops a minimum. For even more extreme mass ratios, a
region at large $\bar{\Lambda}$ becomes invalid because $\Lambda_1<0$.

One could argue that the excluded regions at small $\Lambda$ follow from 
the universal relations, which hold to good accuracy for known nuclear 
physics EOS. However, the EOS used to derive the relations already exclude
even larger $\Lambda$, as discussed earlier. The exact extent 
of the region excluded by negative $\Lambda_1$ is just the result of 
extrapolating a fit and has no physical meaning.

A similar set of universal relations is used in \cite{SoumiDe:2018}.
The set of EOS used for calibration consists of many piecewise 
polytropic EOS, each of which leads to maximum NS masses above $2\,M_\odot$.
For the mass range of the detected event,
the tidal deformability of NS following those EOS is approximately
proportional to $q^6$. This leads to a parametrization
$\Lambda_1 = \Lambda_0 q^3$, $\Lambda_2 =  \Lambda_0 / q^3$, where
$\Lambda_0$ is a free parameter.
In our notation, the slope parameter is $\bar{S}=6$, independent of $\bar{\Lambda}$
and $q$.
The given slope parameter is also shown in \Fref{fig:Sl_vs_Lambda_m}.
The differences to NS models with nuclear physics EOS are comparable to those 
found for the universal relation used in \cite{LVC:EOSPaper:2018}.
In contrast to \cite{LVC:EOSPaper:2018}, the relation used in 
\cite{SoumiDe:2018} does not exhibit any divergence or excluded regions 
at low values of $\Lambda$.

It has become customary to express constraints on the EOS from NS observations 
in a mass-radius diagram. However, the radius does not enter the gravitational waveform
models at all and therefore it cannot be measured.
In \cite{LVC:EOSPaper:2018} and  \cite{SoumiDe:2018}, the NS tidal
deformabilities were therefore converted to radii using another type of ``universal'' relation,
which relates tidal deformability and compactness $C=M/R$ of the NS.

\begin{figure}
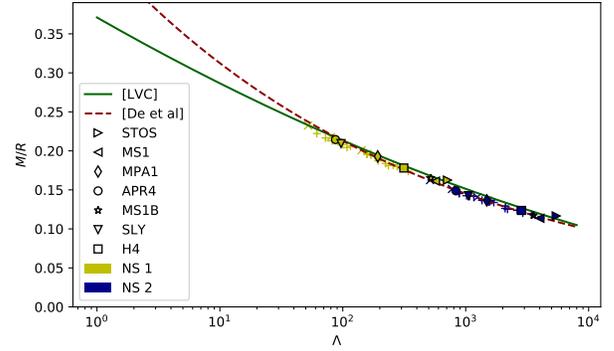

  \begin{center}
    \includegraphics[width=0.95\columnwidth]{{{univ_compactness}}}  
    \caption{Universal relation used in \cite{LVC:EOSPaper:2018}
    for NS compactness as a function of tidal deformability (green curve). 
    The dashed red curve shows a similar universal relation used in 
    \cite{SoumiDe:2018}. The labeled symbols and plus markers 
    show values obtained from various EOS (the crosses correspond to causality-violating EOS)
    for the masses of the lighter (blue) and heavier (yellow) NS in GW170817 
    when assuming a mass ratio $q=0.7$. 
 }
    \label{fig:univ_compactness}
  \end{center}
\end{figure}

The universal $\Lambda-C$ relation used in \cite{LVC:EOSPaper:2018}
is given in \cite{Yagi:2016:13LT01,Chatziioannou:2018:104036}, while 
\cite{SoumiDe:2018} introduces a different one.
Both variants are shown in \Fref{fig:univ_compactness}. 
For comparison, we 
plot values for our standard set of EOS at two masses spanning the range of the detected 
event.
The universal relation $C_u(\Lambda)$ contains the same conceptual trap as the relation 
$\Lambda_a(\bar{\Lambda}, q)$: for a given NS mass, it is only meaningful in the range 
covered by the set of EOS 
used for the fit. Outside this range, none of those models has the correct mass. 
For a NS with given mass and low tidal deformability, one cannot derive any expectation 
for the compactness from nuclear physics EOS if the very same set of EOS firmly excludes 
NS models with those $\Lambda$, $M$.
We note that outside the range supported by NS models, the two relations 
differ significantly, since the relation used in \cite{SoumiDe:2018} was only 
designed to fit the mass range of GW170817.

For each of the two masses shown in \Fref{fig:univ_compactness}, 
the range of EOS spans only a limited range in $\Lambda$. For the example at mass
$1.14 \usk M_\odot$ (blue markers), the universal relation prediction for the compactness becomes
increasingly unreliable below
$\Lambda \lesssim 500$, the lower limit of the range allowed by the EOS set.
This corresponds to a radius of 
${\approx}10\usk\kilo\meter$ 
when using the universal relation. 
NSs of same mass but lower tidal deformability
would require an EOS very different from those used to construct the universal
compactness relation. A possible cause for such qualitatively different behavior 
might involve phase transitions or quark matter, none of which are considered in 
the EOS sets used for calibration in \cite{SoumiDe:2018,LVC:EOSPaper:2018}.

In conclusion, assuming that the EOS of an observed NS falls in the 
range of EOS used to calibrate the universal compactness relations already 
leads to lower limits of radius and tidal deformability based on the measured 
mass range alone.
Inserting any measured tidal deformability below this limit
in the universal relations to obtain a radius is contradictory,
because the existence of such a NS invalidates the foundations of the relation.

A consistent use of universal relations in statistical analysis would require 
specifying a mass-dependent validity range defined by the set of EOS used to calibrate the 
relations, and to limit the prior support to valid regions in parameter space.
Most likely, the resulting lower limits for GW170817 would then be close to 
the model-dependent ones already resulting from the validity region and 
the measured mass range.

\section{Application to data analysis}
\label{sec:implic}

\subsection{Ideal measurements}  
\label{sec:ideal}
In the following we study the constraints given by an exact measurement 
of the waveform. Although real measurements only yield statistical probabilities
due to the noise of the detector, it is instructive to discuss this idealized case 
first, since real measurements can never be more constraining.
We thus consider the case that the inspiral GW signal was 
measured with perfect accuracy and agrees exactly with a waveform model with 
parameters $q,\tilde{\Lambda}$. 
As always, we assume a fixed chirp mass, zero 
spins, and that $\bar{S}$ has negligible impact on the waveform.

The exact measurement of $q,\tilde{\Lambda}$ corresponds to a single point in the 
$(\bar{\Lambda}, \bar{M})$ parameter space shown in \Fref{fig:lambda_m_vs_M_m_all}.
A given EOS is compatible with the measurement if the corresponding function 
$\bar{\Lambda}(\bar{M})$ passes through the point. Note that for some of the nuclear 
physics EOS, the $\bar{\Lambda}(\bar{M})$ curves shown in the upper panel of 
\Fref{fig:lambda_m_vs_M_m_all} cross. 
With a single measurement of a neutron star binary at such an intersection, the 
two EOS would be completely indistinguishable. 
On the other hand, the curves for our set of EOS cross at shallow angles,
such that the $\bar{\Lambda}(\bar{M})$ curves remain similar when moving away from
the intersection point. How large the differences of two EOS away from an intersection 
point can become is a highly relevant open question. 

The similarity between intersecting $\bar{\Lambda}(\bar{M})$ curves does not extend 
to maximum mass NS, however. EOS models with otherwise very similar curves can have very 
different maximum masses. Counterexamples visible in \Fref{fig:lambda_m_vs_M_m_all} 
show that the maximum NS mass cannot be determined from a single inspiral-phase GW signal 
alone, although $M_1$ constitutes a lower limit (assuming slow rotation).

In general, the inverse problem of deducing the constraints on the actual EOS function 
$P(\rho)$ from $\bar{\Lambda}$ 
measured at one or several $\bar{M}$ is a very nontrivial problem, partly because
the full parameter space of $P(\rho)$ has infinite dimension. However, it is not necessary 
to solve this problem for GW data analysis. Since the only information the GW signal
carries is encoded in $\bar{\Lambda}, \bar{M}$, one could restrict signal analysis 
to this information. Testing model assumptions for compatibility or deriving 
information not in the GW signal by adding model assumptions can both be accomplished
in a step separate from GW data analysis.

For an ideal measurement of $\bar{\Lambda}$ at some mass ratio $q$, the universal relations
$\Lambda_a(\Lambda_s, q)$ yield exact values of the individual $\Lambda_1, \Lambda_2$,
unless $\bar{\Lambda}$ is inside one of the excluded regions discussed in \Sref{sec:univ}.
In our notation, the universal relation $\Lambda_a(\Lambda_s, q)$ can be written 
as $\bar{S}_u(\bar{\Lambda}, q)$, with a range $\bar{\Lambda} < \bar{\Lambda}_e(q)$ excluded
by construction.
The accuracy of $\Lambda_a(\Lambda_s, q)$ needs to be specified as part of the model assumptions. 
Since there is only one true EOS and one universal relation, the error is systematic. 
From the viewpoint of measurements, the model assumption is defined by the allowed
range for $\Lambda_a$.
A reasonable range is given by the largest fit residuals of the EOS used for calibrating 
the relations, although it depends on the selection of EOS models, which is part of the assumption.
A comparison of the slope parameter $\bar{S}$ between universal relations and various EOS
is shown in \Fref{fig:Sl_vs_Lambda_m}. As one can see, the EOS we consider fulfill
$4 < -\bar{S} < 9$. This range could serve as a simpler replacement for the use of
universal relations $\Lambda_a(\Lambda_s, q)$ together with comparable accuracy assumptions.

Figure~\ref{fig:Sl_vs_Lambda_m} also leads us to the following  observations.
First, the 
more complicated universal relations used in \cite{LVC:EOSPaper:2018}
do not reduce the residuals considerably compared to the more robust $\bar{S}=6$ prescription
used in \cite{SoumiDe:2018}. Further, the spread for the EOS we considered is somewhat larger
than for the smaller set used to calibrate the universal relations in \cite{LVC:EOSPaper:2018}.
Finally, $\bar{S}_u(\bar{\Lambda},q)$
diverges for $\bar{\Lambda}$ outside the range allowed by the EOS models 
considered here, and the error is essentially unknown. Using the same error
as for $\bar{\Lambda}$ values compatible with the EOS models is not a well-motivated
assumption, given that the divergence is caused by extrapolation of a particular 
form of the fitting function.

\subsection{Bayesian analysis}
\label{sec:bayesanity}
The study of gravitational-wave data relies on Bayesian statistical methods.
The main ingredients are 
\begin{enumerate*}[label=(\roman*)]
\item a theoretical model of the signal 
expected for given source and extrinsic parameters;
\item a characterization of the detector noise that allows one to estimate 
how likely a given stretch of data contains a given signal; and 
\item the prior probability distribution for the source 
parameters, which encodes the state of knowledge before any measurement.
\end{enumerate*}

The choice of a prior is a very ambiguous ingredient, although
there are some well-motivated constraints to start with.
For example, gravitational masses of NS are positive in general relativity.
Similarly, negative tidal deformabilities are usually excluded (see 
\cite{Tianqi:2018:063020}).
Another important decision is whether to assume that the EOS of NS is universal, 
i.e. that both NS follow the same $\Lambda(M)$. 
Further, the use of universal relations is equivalent to a particular choice of prior.

There is however no observational or theoretical knowledge that could motivate
a choice for the exact shape of the tidal deformability prior distribution.
In this situation, it is common to express the lack of knowledge by 
assuming a uniform prior within the bounds. However, this depends on the choice of 
parametrization and is therefore still ambiguous. For a NS, both $\Lambda$ and 
$\log(\Lambda)$ are natural choices. A further complication is
given by the choice of variables for the multidimensional parameter space 
of a binary system, in particular, in connection with universal relations. 
We do not try to argue which prior is ``correct'' but describe
the various priors used in previous studies and their implications in detail.

The studies \cite{LVC:BNSSourceProp:2019,LVC:EOSPaper:2018,SoumiDe:2018} 
provide results on the tidal deformabilities in 
terms of the posterior distributions, which combine the prior with the 
information added by the observational data. The main results are 
one- or two-sided credible intervals on the tidal deformability or 
derived quantities.
In the following sections, we investigate how much of the posterior 
depends on the prior assumptions, and how much information is added by
the observational data. 

With decreasing deformability, its impact on the waveform diminishes.
For a given signal-to-noise ratio, we expect that there is some scale 
$\tilde{\Lambda}_t$ below which the ability of the detector to 
distinguish tidal effects gradually vanishes.
One should therefore expect that any lower limit 
depends more strongly on the prior the smaller its value, and that upper limits
depend less on the prior than lower limits.

With this motivation, we scrutinize the lower credible bounds
that are already given by the priors.
To this end, we introduce a thought experiment we dubbed the 
uninformative detector test.
Consider a detector that measures with perfect accuracy the NS masses and whether 
or not the effective tidal deformability $\tilde{\Lambda}$ is greater than some 
threshold $\tilde{\Lambda}_t$ 
or not, but does not provide the value of $\tilde{\Lambda}$. 
This is an idealized model of a real detector,
for which the ability to distinguish increases gradually instead of abruptly.
The posterior obtained by a hypothetical measurement with the uninformative 
detector is given by the prior, after cutting it off above $\tilde{\Lambda}_t$  and
normalizing it to unity (assuming the true value is below $\tilde{\Lambda}_t$).

Of course one can compute perfectly valid two-sided credible intervals 
$(\tilde{\Lambda}_t^0, \tilde{\Lambda}_t^1)$ for this 
posterior. However, the lower limit $\tilde{\Lambda}_t^0$ is rather useless 
because it is entirely
given by the prior and the threshold $\tilde{\Lambda}_t$. By construction, 
the detector cannot distinguish any deformabilities below $\tilde{\Lambda}_t$, 
including 0. A measurement with such a detector could never be considered as 
observational proof of finite tidal effects (unless the true value is above $\tilde{\Lambda}_t$).

The lower limit $\tilde{\Lambda}_t^0$ can serve as a reasonable estimate for the scale 
below which lower limits are likely not informed by the data, but the prior.
As further motivation consider the likelihood $P(d|\theta,I)$ in \Eref{eq:Bayes_law}.
For the uninformative detector, the likelihood is constant for 
$\tilde{\Lambda}<\tilde{\Lambda}_t$ and 0 otherwise. 
Next, consider an actual detection with a different likelihood function that is 
featureless at scales $\tilde{\Lambda}_t^0$.
In order to obtain $\tilde{\Lambda}_m^0 < \tilde{\Lambda}_t^0$ for the 
actual detection,
the corresponding likelihood below $\tilde{\Lambda}_t^0$ would need to
\emph{increase} on average, compared to the uninformative case, or decrease
above $\tilde{\Lambda}_t^0$. Hence low values become more likely by such 
a measurement, and a two-sided credible interval is probably not meaningful.

In order to provide $\tilde{\Lambda}_t^0$ for GW170817, we still need an estimate for 
$\tilde{\Lambda}_t$. 
Here we choose $\tilde{\Lambda}_t \approx 800$, motivated by the posterior shown 
in Fig.~11 of \cite{LVC:BNSSourceProp:2019}, which falls off strongly at larger values, 
indicating that tidal effects of this size can in fact be distinguished. 
Note the scale $\tilde{\Lambda}_t$ could also
be lower, which would imply a lower $\tilde{\Lambda}_t^0$. Since we use 
$\tilde{\Lambda}_t^0$ as a conservative
sanity check for the interpretation of lower limits, this is not a problem. 
The case $\tilde{\Lambda}_m^0 \le \tilde{\Lambda}_t^0$ can be regarded as a strong 
indication, but not conclusive evidence, that $\tilde{\Lambda}_m^0$ is mainly a 
consequence of the prior assumptions. In this case, providing credible intervals 
is insufficient and a detailed comparison of posterior and prior shape is necessary.

\subsection{Universal relations in Bayesian analysis}
\label{sec:real}
We now turn to discuss the use of universal relations in Bayesian analysis
of GW signals. To focus on the essential problems, we consider the simplified case
where chirp mass and mass ratio are known exactly, but not the tidal effects.

\begin{figure}
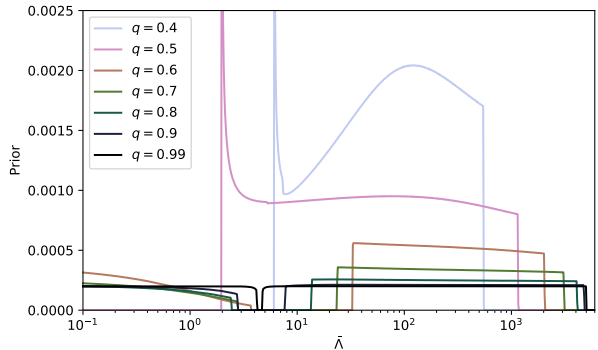

  \begin{center}
    \includegraphics[width=0.95\columnwidth]{{{Prior_univ_Lambda_m}}}  
    \caption{Prior distribution of $\bar{\Lambda}$ obtained using the universal relation 
    $\Lambda_a(\Lambda_s, q)$ with a flat prior for $\Lambda_s$ in the range $(0,5000)$, 
    fixed values for the mass ratio, and the given chirp mass. The logarithmic x-axis 
    allows one to see the gaps in the distribution. Note however that the curves show the 
    probability density of $\bar{\Lambda}$, not $\log(\bar{\Lambda})$. 
    }
    \label{fig:Prior_univ_Lambda_m}
  \end{center}
\end{figure}

In \cite{LVC:EOSPaper:2018}, the tidal effects are encoded in one 
parameter $\Lambda_s$, for which a flat prior distribution in the range $(0,5000)$ is 
assumed, independent of chirp mass and mass ratio.
The prior for the quantity that is actually measured, $\bar{\Lambda}$, is then derived
from the universal relations. The result for the given chirp mass and various mass 
ratios is shown in \Fref{fig:Prior_univ_Lambda_m}. As already mentioned, the 
universal relations have poles that cause gaps and peaks at very small values of
$\bar{\Lambda}$. More importantly, the strict prescription of the slope causes
gaps extending to larger values of $\bar{\Lambda}$ at certain mass ratios, 
because otherwise $\Lambda_1$ becomes negative. 
This makes the analysis unsuitable for providing lower limits on $\bar{\Lambda}$ based
on measurement. 

Further, the upper limit for $\bar{\Lambda}$ depends on the mass ratio. The prior 
density at fixed $\bar{\Lambda}$ is therefore reduced with increasing mass ratio,
as shown in \Fref{fig:Prior_univ_Lambda_m} (the prior for $\tilde{\Lambda}$ shows
a similar behavior).
This is not a physical assumption, but an artifact of the functional form of the universal
relation in conjunction with the choice of a flat prior with constant cutoff for $\Lambda_s$. 
For given $\bar{\Lambda}$ or $\tilde{\Lambda}$, the prior used in \cite{LVC:EOSPaper:2018} 
favors unequal mass systems, independent of the chirp mass.
Therefore we expect this issue
to cause a bias away from the equal mass case.
We note that the prior for the masses itself already favors unequal mass systems,
again not due to physical considerations but due to fixed limits prescribed for the 
chirp mass.

We conclude that the use of universal relations in Bayesian analysis 
---in contrast to an ideal measurement discussed in \Sref{sec:ideal}---
can also affect the measured quantity $\bar{\Lambda}$, via the prior.
However, the prior used in \cite{SoumiDe:2018} does not exhibit this problem,
thanks to an additional cutoff in $\tilde{\Lambda}$ and a simpler choice of
universal relations. 
The prior for $\tilde{\Lambda}$ from \cite{SoumiDe:2018} is flat, 
without gaps, and independent of the mass ratio.

The universal relations are used in \cite{LVC:EOSPaper:2018} to derive posteriors
for $\Lambda_1$ and $\Lambda_2$. We note that, for given $q$, $\Lambda_1$ and $\Lambda_2$ are
uniquely determined by the parameters $\bar{\Lambda}$ and $\bar{S}$. While $\bar{\Lambda}$
enters the GW waveform model, $\bar{S}$ does not and can therefore not be measured 
at all. All information on $\bar{S}$ is derived from the universal relations. We
thus treat it as a statistical variable derived from $\bar{\Lambda}$ based on a 
model (the universal relations).

\begin{figure}
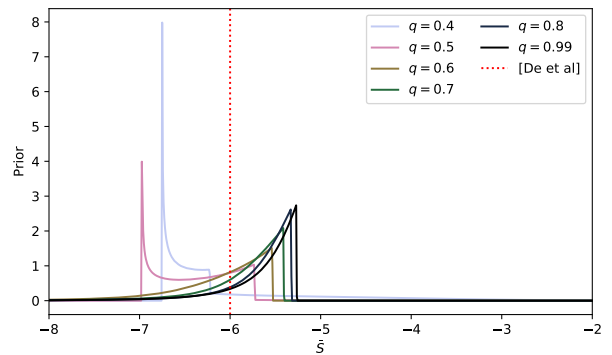

  \begin{center}
    \includegraphics[width=0.95\columnwidth]{{{Prior_univ_Sl}}}  
    \caption{Like \Fref{fig:Prior_univ_Lambda_m}, but showing the prior
    distribution for the slope $\bar{S}$. In addition, the vertical line represent
    the unique slope given by the universal relations 
    used in \cite{SoumiDe:2018}.}
    \label{fig:Prior_univ_Sl}
  \end{center}
\end{figure}

The resulting prior distribution for $\bar{S}$ is shown in \Fref{fig:Prior_univ_Sl}. 
For any mass ratio, the distributions are strongly peaked and asymmetric. The 
shape of the prior is given entirely by the functional form of the universal relation.
It is not related to the distribution of $\bar{S}$ resulting for the EOS 
models (which would not be meaningful anyway since the EOS selection is arbitrary).
For the universal relations in \cite{SoumiDe:2018}, the prior for 
$\bar{S}$ is even more restrictive, given by a delta function, which is shown 
as a vertical line in \Fref{fig:Prior_univ_Sl}.

We note that the prior resulting from the universal relations is not used
directly in \cite{LVC:EOSPaper:2018}. Instead, the prior for $\Lambda_a$
is broadened in order to account for the error of the universal 
relations, 
which is estimated from the residuals of the
EOS set used for calibration. The broadening of $\Lambda_a$ corresponds
mainly to a broadening of $\bar{S}$, but to some extent also of $\bar{\Lambda}$.
The sharp features visible in Figs.~\ref{fig:Prior_univ_Lambda_m} 
and~\ref{fig:Prior_univ_Sl} are smoothed in the final prior used in 
\cite{LVC:EOSPaper:2018}, and the gaps in the prior of $\bar{\Lambda}$ 
gain some partial support. 
This might alleviate, but not solve, the 
issues discussed here. 
For the rest of this article, we therefore ignore
the effects of the additional broadening.

We also note that deriving the prior of $\bar{S}$ from a flat
prior of $\Lambda_a$ and the universal relations and then broadening it to account for 
errors of the latter is not the only possible approach. 
Further, \cite{LVC:EOSPaper:2018} uses error estimates based
on treating an arbitrary selection of EOS models as if it were statistical sampling 
of the true EOS, arriving at a complicated model for the error that depends on 
$\Lambda_s$ and $q$.
A cleaner and simpler approach would be to directly prescribe a prior in $\bar{S}$,  
e.g., a Gaussian, with mean value and broadness defining the model assumptions.
That way, one retains control over the shape of the prior distribution and considerably
simplifies the interpretation of the assumptions.

\begin{figure}
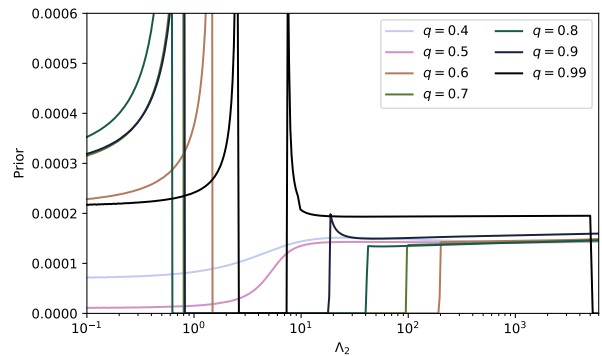

  \begin{center}
    \includegraphics[width=0.95\columnwidth]{{{Prior_univ_Lambda2}}}  
    \caption{Like \Fref{fig:Prior_univ_Lambda_m}, but showing the prior
    distribution for the tidal deformability  $\Lambda_2$ of the lighter NS.
    Although the x-axis is logarithmic, the curves show the 
    probability density of $\Lambda_2$, not $\log(\Lambda_2)$ }
    \label{fig:Prior_univ_Lambda2}
  \end{center}
\end{figure}

The gaps in the priors of $\bar{\Lambda}$ used in \cite{LVC:EOSPaper:2018}
in conjunction with the slope $\bar{S}$ 
restricted by the universal relations
have an important consequence for the prior of $\Lambda_2$, which is shown
in \Fref{fig:Prior_univ_Lambda2}.
As one can see, it has even larger gaps than the prior for $\bar{\Lambda}$.
For mass ratio $0.7$, the gap extends up to $\Lambda_2 \approx 100$.
The method is therefore particularly unsuitable to measure lower limits on $\Lambda_2$.

\begin{figure}
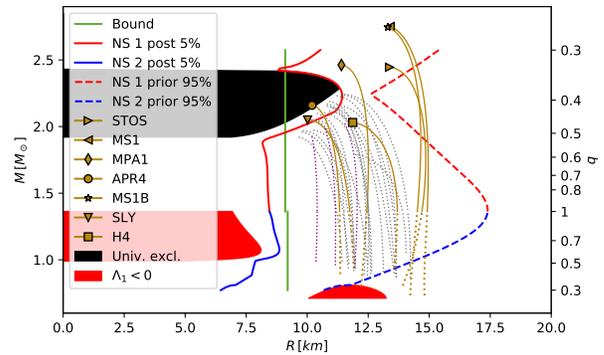

  \begin{center}
    \includegraphics[width=0.95\columnwidth]{{{univ_M_R}}}  
    \caption{The solid blue and red curves show the fifth percentile of the NS radii
    derived from the prior used in \cite{LVC:EOSPaper:2018},
    restricted to $\tilde{\Lambda}<800$.
    Those are not boundaries of a two-dimensional distribution; rather the percentiles are 
    computed for each mass ratio $q$ and plotted versus $M_1(q)$ (blue) 
    and $M_2(q)$ (red).
    The dashed blue and red curves show the 95th percentiles of
    the full prior.
    The black and red areas correspond to values excluded by the universal
    relations in \cite{LVC:EOSPaper:2018}. The lower credible bounds inferred 
    in \cite{LVC:EOSPaper:2018} for the radii of the two NS using the universal 
    relations are shown as two vertical green lines.
    For comparison, we also show the mass-radius diagrams for a representative set
    of nuclear physics EOS models.
    The curves are cut at the mass of the lighter NS in a binary 
    where the heavier NS mass is maximal (with our example chirp mass).}
    \label{fig:univ_M_R}
  \end{center}
\end{figure}

We now apply the uninformative detector test introduced in \Sref{sec:bayesanity}
to the prior used in \cite{LVC:EOSPaper:2018}. 
As in \cite{LVC:EOSPaper:2018}, we use a flat prior in $\Lambda_s$, 
cut off at $\Lambda_s=5000$. 
Following \cite{LVC:EOSPaper:2018}, we use the universal relations 
$\Lambda_a(\Lambda_s, q)$ to compute the $\Lambda_i$ of 
the two NS, and further the radii $R_i = M_i / C_u(\Lambda_i)$ (we do
not apply the additional broadening of the prior, though).
We exclude samples from the prior that violate $\Lambda_2 \geq \Lambda_1 > 0$.
The fifth percentiles of the hypothetical posteriors of the radii that follow 
from the uninformative detector test are shown in 
\Fref{fig:univ_M_R} for a range of mass ratios, as a function of $M_i(q)$.
As one can see, the radii are still as large as 
${\approx}8\usk \kilo\meter$ in the mass range of interest.

What is the reason for those surprisingly large bounds from our uninformative 
hypothetical measurement? 
One cause is the form of the universal relation $C_u(\Lambda)$, 
which is a second order polynomial in $\log(\Lambda)$. Considering any flat 
distribution in $\Lambda$, it is not surprising that the 5th/95th 
percentiles cover only a small range of radii. 
To some extent, the lower limits are also caused by the functional form of 
the universal relation $\Lambda_a(\Lambda_s, q)$, which completely
excludes tidal deformabilites below a certain limit (depending strongly on the mass ratio).
Translating this exclusion zone to the radius of the lighter NS via $C_u(\Lambda)$,
we find that radii
up to $8\usk\kilo\meter$ are already excluded by construction. This is shown in
\Fref{fig:univ_M_R} as well. It also shows that for extreme mass ratios, further regions in 
parameter space become unreachable, which is not relevant for the detected event,
however, given that the mass ratio is constrained.

The lower limits published in \cite{LVC:EOSPaper:2018} for the measured
posteriors are shown as well in \Fref{fig:univ_M_R}. As one can see,
the limits are quite similar to those obtained with our hypothetical
detector. Comparing the radii derived from the universal relations is 
not ideal to assess how much the observational data contributed to the 
posterior. It would be easier to interpret results for the quantity that
directly affects the waveform, $\tilde{\Lambda}$. 
Although \cite{LVC:EOSPaper:2018} do not quote limits for $\tilde{\Lambda}$,
they provide 90\% two-sided credible intervals $\Lambda_{1.4}=190^{+390}_{-120}$
for the deformability of a NS with mass $M=1.4 M_\odot$, obtained by assuming
a linear expansion of $\Lambda(M)M^5$ around $M=1.4 M_\odot$.
For comparison, we computed the fifth percentiles of $\bar{\Lambda}$, which increase 
from $44$ at mass ratio $q=1$ to $58$ at $q=0.7$, and $\bar{\Lambda}\approx50$
at $\bar{M}=1.4 M_\odot$.
The limits from the prior are roughly comparable to the published lower credible 
bound of $70$.
\emph{Our findings indicate that the lower radius limits, as obtained in 
\cite{LVC:EOSPaper:2018} using the universal relations prior, might be largely 
based on the prior assumptions and less on the observational data.}
We stress that our qualitative discussion is not sufficient to entirely 
dismiss the contribution of the data to the lower bounds. 
Nevertheless, the results denoted in \cite{LVC:EOSPaper:2018} as measurements 
of lower limits for the radii should not be interpreted as observational evidence 
for finite size effects yet, even under the assumption that both stars were 
neutron stars following the same EOS.

A more basic test than our uninformative detector is to visually
compare the prior and posterior distributions.
Such a comparison is shown in the top-left panel of Fig. 3 of
\cite{LVC:EOSPaper:2018}.
The prior for the radius of the lighter NS is much less restrictive
than the posterior, also towards small radii.
This might be interpreted as confirmation that the lower bound
is indeed based on data, contrary to our previous discussion. 
However, there is a subtle but profound
pitfall: the priors for the radii shown in \cite{LVC:EOSPaper:2018} 
are not restricted to the mass range where the posterior has support. 
In particular, the full prior contains a range of chirp masses.
The mass enters directly in the inferred radius when
using the compactness-deformability universal relation.
In other words, any measurement of the mass already restricts the radius 
range, even without measuring the deformability.

Comparing the radius posterior to the full prior is therefore not a valid
test to assess if the measurement of the deformability is informative.
Showing only the full prior of the radii obfuscates the fact that
the priors restricted to the almost exactly 
measured chirp mass are much more restrictive for the inferred radius.
In order to judge if the lower credible bound on the radius 
corresponds to a measurement of finite tidal effects or merely
reflects a measurement of the mass range in conjunction with the 
model assumptions,
one should instead compare the radius posterior to the radius prior 
restricted to measured chirp mass and mass ratio range.
A very similar problem was already discussed in the context of spin precession 
in binary black hole mergers \cite{LIGO_GW151226,LIGO:2016:041015}.

We carried out the same analysis also for the universal relations
from \cite{SoumiDe:2018}. Similar to the previous discussion,
we consider fixed masses and a flat prior
for the overall tidal deformability, this time in terms of the parameter 
$\Lambda_0$ used in \cite{SoumiDe:2018}, with the same cutoff at
$\Lambda_0=5000$. Since \cite{SoumiDe:2018} remove tidal deformabilities 
violating limits from causality considerations \cite{Tianqi:2018:063020},
we apply a cut $\tilde{\Lambda} \le 51$, derived in \cite{Tianqi:2018:063020}
for the chirp mass of GW170817 and assuming a maximum NS mass $\ge 2 M_\odot$.

\begin{figure}
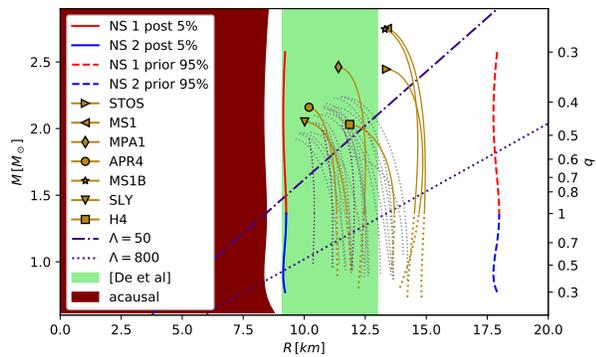

  \begin{center}
    \includegraphics[width=0.95\columnwidth]{{{De_univ_M_R}}}  
    \caption{Like \Fref{fig:univ_M_R}, but for the universal relations
    from \cite{SoumiDe:2018}. 
    The green shaded area marks the statistical bounds reported for the ``common radius''
    in \cite{SoumiDe:2018}, not including the systematic errors, and based on a prior
    that excludes tidal deformabilities below the causality limit.
    The brown shaded area shows causality-violating region obtained by combining the bound
    $\tilde{\Lambda} \le 51$ derived in \cite{Tianqi:2018:063020} and the universal relations
    from \cite{SoumiDe:2018}.
    For comparison, we also show the radii derived using $C_u$ (the version from
    \cite{SoumiDe:2018}) for fixed values of $\Lambda$.
    }
    \label{fig:De_univ_M_R}
  \end{center}
\end{figure}

As before, we apply our uninformative detector test
to obtain a hypothetical posterior. We find that the 
fifth percentile for the radii are very close to the lower bounds 
inferred in \cite{SoumiDe:2018} for the actual detection. 
This is shown in \Fref{fig:De_univ_M_R}, together with the 
limits obtained for causal neutron-star models.

Figure~3 in \cite{SoumiDe:2018} shows the posterior of the effective 
tidal deformability. The distribution
drops sharply to 0 a bit below the lower credible bound. Note,
however, that low values of $\tilde{\Lambda}$ incompatible with causality 
have been removed from the posterior shown in \cite{SoumiDe:2018},
which likely causes the drop. The limit $\tilde{\Lambda} \le 51$ from 
\cite{Tianqi:2018:063020} is comparable to the size of the gap.
The original posterior and the prior 
are not shown in comparison, but \cite{SoumiDe:2018} provide
limits both before and after removing causality-violating points from 
the distribution. The removal increases the lower bounds for $\tilde{\Lambda}$ 
by ${\approx}40$. When using a prior without
the causality cut, both the uninformative detector and the analysis of actual data 
in \cite{SoumiDe:2018} result in lower limits reduced by a similar amount.

Our findings indicate that the lower credible limits inferred by \cite{SoumiDe:2018} 
are almost entirely due to the prior assumptions, while the measurement did not add
significant support for finite tidal deformabilities. 
This calls into question the claim of ``first evidence for finite size effects using 
gravitational-wave observations'' \cite{SoumiDe:2018}.
Although \cite{SoumiDe:2018} compute Bayes factors around 10 in disfavor of 
$\tilde{\Lambda}<100$ compared to the full range,
an erratum \cite{SoumiDe:2018:Erratum} reports that the employed thermodynamic integration is 
unsuitable for computing Bayes factors. The impact on the Bayes factor discussed 
above is not provided, however.

Besides the problems mentioned before, we also find that the lower limits on radii 
or tidal deformabilites published in \cite{SoumiDe:2018,LVC:EOSPaper:2018}
are well below the validity range of the universal relations used in their
deviation. Figures~\ref{fig:univ_M_R} and~\ref{fig:De_univ_M_R} show the 
limits in comparison to radii obtained for a number of EOS, including those used to calibrate the 
relations used in \cite{LVC:EOSPaper:2018}. 
The lower bounds for the radii are the result of evaluating phenomenological expressions 
in a range explicitly forbidden by the models used to derive the very same 
expressions (see \Sref{sec:univ}).

\subsection{EOS-agnostic Bayesian analysis}
\label{sec:agnostic}
In this section, we discuss the analysis of tidal effects in \cite{LVC:BNSSourceProp:2019},
with a focus on the interpretation of lower limits. 
As in previous  sections, we use the idealized case of known masses and zero spins to
highlight problems that are not obvious in the full statistical analysis.

In \cite{LVC:BNSSourceProp:2019} it is not assumed that both stars obey the same EOS, or that they
are neutron stars, and no universal relations are employed.
In detail, \cite{LVC:BNSSourceProp:2019} employs a flat prior
for the tidal deformabilities $\Lambda_1, \Lambda_2$, cut at a maximum value
$0 \le \Lambda_1,\Lambda_2<5000$. This implies that the tidal deformabilities
are not only allowed to differ for equal mass systems, but also that they are 
completely uncorrelated, i.e. it is not assumed that stars with similar mass 
are more likely to have similar deformabilities.

\begin{figure}
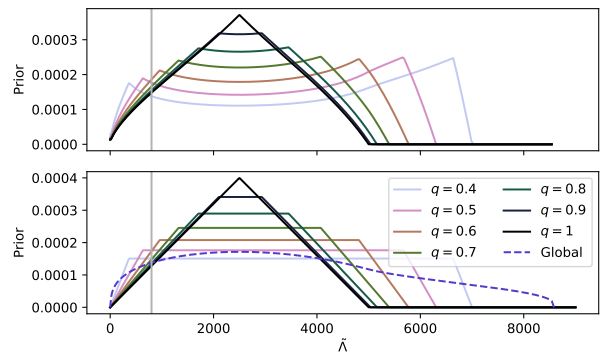

  \begin{center}
    \includegraphics[width=0.95\columnwidth]{{{Prior_flat12_Lambda_eff}}}  
    \caption{Bottom: Prior distribution of effective tidal deformability $\tilde{\Lambda}$
    derived from the EOS-agnostic flat $\Lambda_1, \Lambda_2$ prior. 
    The dashed curve shows the full prior marginalized over all mass ratios;
    the other curves show the prior restricted to various mass ratios.
    Top: modified prior distribution for $\tilde{\Lambda}$ obtained by dividing the 
    original prior by the prior after marginalizing out the mass ratio.
    The different curves show the modified prior restricted to the same mass ratios
    as the bottom panel. We note the prior density at $\tilde{\Lambda}=0$ is small, 
    but not 0.
    }
    \label{fig:Prior_flat12_Lambda_eff}
  \end{center}
\end{figure}

The assumptions of uncorrelated deformabilities have an important consequence
for the prior of $\tilde{\Lambda}$. Because of the rectangular uniform support in
$(\Lambda_1, \Lambda_2)$ space, 
it linearly approaches 0 near $\tilde{\Lambda}=0$,
with a slope that depends on the mass ratio.
Figure~\ref{fig:Prior_flat12_Lambda_eff} shows the prior resulting for $\tilde{\Lambda}$
restricted to various mass ratios.
Small values of $\tilde{\Lambda}$ are already disfavored by the prior distribution. 
Obviously, this needs to be considered before interpreting lower credible bounds
as a proof of finite size effects.

In \cite{LVC:BNSSourceProp:2019}, the raw posterior of $\tilde{\Lambda}$ is divided 
by the marginalized prior of $\tilde{\Lambda}$ before computing credible intervals 
on $\tilde{\Lambda}$. 
This approach effectively corresponds to using a prior in $\tilde{\Lambda}$
that is flat after marginalization over mass ratio.
The motivation given in \cite{LVC:BNSSourceProp:2019} is to address the issue above, 
although the physical motivation for the modified prior is
not explained. Since the behavior of the 
original prior is a direct consequence of the assumption of uncorrelated deformabilities,
scaling it to obtain a flat prior in $\tilde{\Lambda}$ implies that the posterior
obtained for $\tilde{\Lambda}$ is based on contrary assumptions.
We note that the new prior does not correspond to the assumption of a common EOS, since 
the NSs can still have the same masses but different deformabilities.
However, in this work we are not arguing what prior should be used, but how much
the ones used determine the results. 

First, we investigate more carefully the technical implications of the rescaling.
One problem is that scaling in a mass ratio independent way is not the only choice to
obtain a flat marginalized prior. This already introduces an ambiguity that has 
not been discussed.
More importantly, the resulting prior restricted to mass ratios relevant for the 
detected event is not flat at all. The reason is that the prior for the mass ratio 
strongly favors unequal mass systems (mainly due to the limits imposed on the chirp
mass) for which the prior of $\tilde{\Lambda}$ has increasingly large slope near 
$\tilde{\Lambda}=0$. 
For moderate mass ratios, the reweighted prior used in  \cite{LVC:BNSSourceProp:2019} 
disfavors low values of $\tilde{\Lambda}$, despite the reweighting procedure.
This is shown in the top panel of \Fref{fig:Prior_flat12_Lambda_eff}.

To asses the contribution of the prior to the lower limits
we now apply the uninformative detector test introduced in \Sref{sec:real}
to the reweighted prior used in \cite{LVC:BNSSourceProp:2019}.
We obtain a lower fifth percentile of
$\tilde{\Lambda}=128$ for the posterior, for all mass ratios $q \gtrsim 0.6$.
This should be compared to the fifth percentile of the (reweighted) posterior inferred
from the actual data in \cite{LVC:BNSSourceProp:2019} for the low-spin prior, 
which is $\tilde{\Lambda}=110$ (the 90\% credible bound of the mass ratio 
is $q > 0.73$). Under the assumption of low spins, \cite{LVC:BNSSourceProp:2019} 
interprets the credible lower bound as evidence for finite tidal effects. 
In contrast, our discussion strongly indicates that  
\emph{the lower bound should be regarded mainly as a consequence of prior assumptions
in conjunction with the observational constraints on the mass ratio}.

In the remainder of this section, we compare the EOS agnostic approach to the
studies based on universal relations.
Surprisingly, the results in \cite{LVC:BNSSourceProp:2019} can be reinterpreted
under the assumption that both objects are neutron stars obeying the same EOS,
in the sense that the tidal deformability is a unique but unknown function of mass.
For this, one only has to exclude systems with mass ratio exactly 1,
which makes no difference because this set has infinitesimal support in parameter space.
Further, one has to allow any functional form of the tidal deformability $\Lambda(M)$, 
including arbitrary steep gradients. In our notation,
this implies an unbound range $-\infty \le \bar{S} \le \infty$.
Note all EOS considered here fulfill $\bar{S}<0$.

\begin{figure}
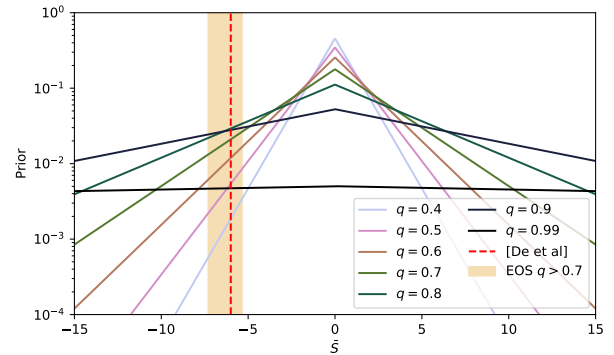

  \begin{center}
    \includegraphics[width=0.95\columnwidth]{{{Prior_flat12_Sl}}}  
    \caption{Prior distribution of the slope parameter $\bar{S}$ resulting from a flat 
    prior in $\Lambda_1,\Lambda_2$ space, restricted to $0\le \Lambda_i \le 5000$, as used 
    in \cite{LVC:BNSSourceProp:2019}.
    We show priors for various fixed mass ratios. Although we only show a restricted range,
    the prior support extends to $\pm\infty$. 
    The shaded band marks the range spanned by our standard EOS set for mass ranges $q>0.7$.
    The dashed line marks the value $\bar{S}$ corresponding to the universal relations
    used in \cite{SoumiDe:2018}.
    }
    \label{fig:Prior_flat12_S_m}
  \end{center}
\end{figure}

Figure~\ref{fig:Prior_flat12_S_m} illustrates the the prior distribution of the slope 
$\bar{S}$ for fixed chirp mass and various fixed mass ratios.
As expected, the support increases towards positively and negatively infinite 
slopes when approaching equal mass. We recall that \Eref{eq:def_sl} then implies
$\mathrm{d}\Lambda / \mathrm{d}M \to \pm \infty$.
At the same time, the prior probability for any finite
slope range approaches 0 in the equal mass limit. 
For unequal masses, the prior support for a given range can also increase with
increasing $q$. In any case, there is a strong dependence on mass ratio.

Since the signal is (almost) independent of the 
slope, the slope posterior is given by the prior. Compared to a slope prior 
bounded to some model-dependent range, the posterior for quantities derived 
from the slope, such as the individual tidal deformabilities, is broadened.
Compared to a more plausible prior $\bar{S}<0$, on can also expect
a strong bias of $\Lambda_1$ and $\Lambda_2$ towards larger and smaller values, 
respectively. 

For comparison, we also plot the range of $\bar{S}$ obtained from our standard set of EOS.
Comparing also to \Fref{fig:Prior_univ_Sl}, one can see that the EOS-agnostic study and
the studies based on universal relations occupy opposing ends of a spectrum of assumptions.
The EOS-agnostic study is overly inclusive regarding the slope, 
compared to the range of theoretical EOS models, while the 
universal relations strongly restrict the slope.
The most restricted slope is given by the fixed value used in \cite{SoumiDe:2018},
while \cite{LVC:EOSPaper:2018} further broadens the raw prior distribution resulting
from their universal relations to account for a certain amount of systematic error.
Given that the universal relation assumptions in \cite{LVC:EOSPaper:2018} strongly 
restrain $\bar{S}$, it is hardly surprising that the credible regions reported occupy 
a smaller volume in parameter space.

\subsection{Parametrized EOS approach}
\label{sec:parameos}

A complementary approach \cite{LVC:EOSPaper:2018} to EOS-agnostic and universal 
relation methods is based on parametrizing the EOS and including 
the corresponding parameters in the statistical analysis. This approach is also slightly
problematic, as we explain in the following.
Imagine an observation where chirp mass and mass ratio are measured exactly, but for
$\bar{\Lambda}$ only a probability distribution is known. Further, consider
a hypothetical EOS parametrization where one parameter $P_u$ uniquely determines $\bar{\Lambda}$ 
for the measured masses,
while the remaining parameters $P_i$ have no influence on $\bar{\Lambda}$, but do have an impact on 
the individual NS properties. 
Since the parameters $P_i$ have no influence on the measured quantity, their 
posterior distribution is given by the prior assumed for $P_i$. 
Therefore, the posteriors for individual NS properties depend on the priors
for $P_i$.
For the parametrization 
used in \cite{LVC:EOSPaper:2018}, the parameters are not as cleanly separated, 
which makes it difficult to assess the importance of the above. 
Note that the implicitly chosen prior in the parameters $P_i$ takes over the role of
the analytic prescriptions for $S_m$ in the universal relation method when
it comes to inferring individual NS properties. 
Knowledge of the priors for a parametrization $P_i, P_u$ instead of the ones 
reported in \cite{LVC:EOSPaper:2018}
is required to interpret the significance of single-NS tidal deformabilities 
and radii inferred with the parametrized EOS method.

Since there is no natural choice for the exact parametrization and no 
other experimental data were used, the shape 
of the prior distribution should be regarded as an arbitrary choice.
It is therefore important to assess how much the prior contributes to any results
derived from the posterior. 

The prior and posterior
for the radii are compared in the top-right panel of Fig.~3 in \cite{LVC:EOSPaper:2018}.
The credible bounds for the prior distribution are not provided.
Visually, it seems that the prior already has little support below the published 
lower limits of the posterior. In addition, the comparison is complicated by 
the fact that the full prior is shown, not restricted to the measured mass range
(compare the discussion in \Sref{sec:real}).
It is therefore impossible to judge how much of the lower limit is based on a 
measurement of the masses in conjunction with the prescribed prior for the EOS 
parameters, and how much on an actual imprint of tidal effects in the data.

We note that instead of parametrizing the EOS, one could directly parametrize the 
function $\Lambda(M)$, for example, using a polynomial. One can then
perform Bayesian parameter estimation of the corresponding coefficients.
This approach was suggested in an earlier work \cite{Agathos:2015:023012}.
When considering a single ideal measurement, an obvious problem arises.
As discussed earlier, a single ideal detection provides one data point 
$\bar{\Lambda}, \bar{M}$.  This single point cannot determine more
than one coefficient, or combination of coefficients. 
For example, if $q\approx 1$, then $\bar{\Lambda}$
determines $\Lambda(M)$ at $M=\bar{M}$. The slope of
$\Lambda(M)$ at $M=\bar{M}$ on the other hand depends on $\bar{S}$, which cannot be measured.
Higher derivatives also have no impact on the measured signal. For unequal 
mass systems, the system is equally underdetermined and one can only measure
one combination of coefficients. 
Bayesian parameter estimation nevertheless yields posteriors
for all the coefficients, and some might appear to be constrained by measurement.
However, the signal still only carries one degree of freedom regarding the coefficients.
We recall that the shape of the posterior distribution in $\bar{\Lambda},\bar{M}$ 
around the true value has nothing to do with the behavior of the $\Lambda(M)$
curve, but rather reflects the detector noise and the relative impact of mass ratio
and deformability. It cannot be used to constrain the slope of $\Lambda(M)$, for example.

The approach is also instructive regarding the parametrized EOS method, which in essence
also contains a parametrization of $\Lambda(M)$, but adds further degrees of freedom
that leave the curve $\Lambda(M)$ invariant. We note, however, that the requirement 
of causality imposes constraints on the EOS. Assuming there is only 
one branch of solutions, and not a transition to quark stars, for example, 
this implies constraints on $\Lambda(M)$. Results about $\Lambda(M)$
that have no impact on the information in the signal, such as $\bar{S}$,
are not completely arbitrary. It is however very difficult to disentangle the meaningful
constraints from the results of Bayesian analysis employing the parametrized EOS.

\subsection{Systematic errors of EOS constraints}
\label{sec:syserr}

In the following, we discuss how systematic errors of the models
for gravitational signals affect constraints of the tidal deformability 
that would result from a perfect detection (in the sense of arbitrary large 
signal-to-noise ratio) of the inspiral phase.
Naturally, the impact of waveform systematics in statistical analysis of 
actual measurements with noisy data cannot be any smaller.

In the sensitive frequency range of current detectors, gravitational waveforms of 
coalescing NSs can only be computed using semianalytic approximations. 
Although some models are calibrated with numerical relativity simulations,
we note that such simulations cover only the upper end of the sensitive 
frequency range of current detectors (in contrast to the BBH case, where the 
merger typically occurs at lower frequencies).

Currently, the error of these models can only be estimated by comparing
waveforms based on different approximations. 
In Bayesian statistical data analysis using matched filtering,
the systematic error of the waveform model causes a corresponding 
bias of the posterior probability distributions.
It is worth pointing out that the bias of tidal effects
is not simply determined by the accuracy with which the tidal effects 
are modeled, but is also sensitive to the accuracy of the underlying
waveform model for zero deformability (binary black hole case).

A central measure in GW analysis is the mismatch between waveforms.
The likelihood $P(d|\theta,I)$ in Eq.~(\ref{eq:Bayes_law})
depends directly on the mismatch between waveform model and data.
It expresses how much waveforms differ from the perspective of a given
detector, weighting different frequency components by detector sensitivity:
\begin{align}
F(h_1, h_2) 
&= 1 - \mathop{\max}_{\phi_c, t_c} \frac{(h_1^{\phi_c, t_c}, h_2)}{\sqrt{(h_1,h_1)(h_2,h_2)}}
\end{align} 
where $h_1^{\phi_c, t_c}$ denotes waveform $h_1$ shifted in time and phase by 
$t_c$ and $\phi_c$, and
\begin{align}
(h_1, h_2) 
&= 4\Re \left(\int_{f_\mathrm{min}}^{f_\mathrm{max}} 
        \frac{\tilde{h}_1(f) \tilde{h}^*_2(f)}{S_n(f)} \mathrm{d}f \right),  
\end{align}
where a tilde denotes Fourier transform and $S_n$ is the detector 
spectral noise density. 
In the following we use the advanced LIGO zero-detuned high-power 
design sensitivity curve (provided by the \textsc{LALSuite} 
\cite{lalsuite} software library, see also \cite{noise_curve}) and apply a high-frequency
cutoff at $f_\text{max} = 1 \usk\kilo\hertz$ when computing waveform mismatches.

Although we discuss the case of arbitrary loud signals, the results in this section 
are still specific to LIGO-type detectors. They do not depend on the overall 
sensitivity though, only the relative sensitivity at different frequencies.
Further, the choice of cutoff parameter does have a varying impact on the
estimates we will present. Since the cutoff frequency
in \cite{LVC:BNSSourceProp:2019} is chosen dynamically using heuristic
criteria based on masses and tidal parameters, the exact impact of the
waveform systematics might be somewhat larger or smaller than our estimates for
fixed cutoff. We will not address this additional complication here, but 
note that it deserves consideration before claiming observational evidence
for tidal effects.

\begin{figure}
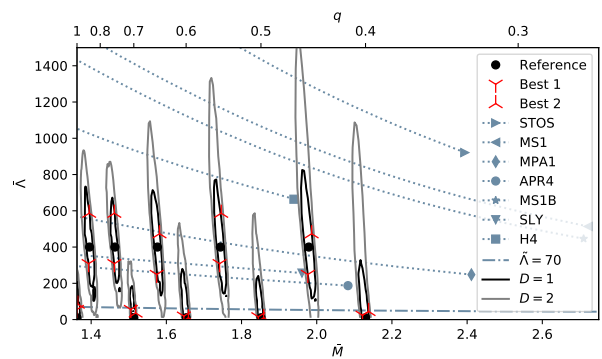

  \begin{center}
    \includegraphics[width=0.95\columnwidth]{{{Lambda_M_mismatch}}}  
    \caption{Systematic error of $\bar{\Lambda}$, $\bar{M}$ due to BNS waveform modeling,
    with respect to selected reference points (black circles).
    The error is estimated with two measures. 
    The contours show $D=1$ (black) and 
    $D=2$ (grey) according to \Eref{eq:err_dist} based on comparing
    IMRPhenomD\_NRTidal and TaylorF2(6PN) waveform models. 
    The markers labeled ``Best 1'' and ``Best 2'' show the location of the best match
    for each of the models to the other one evaluated at the reference point.
    For comparison, we also show the curves obtained for different EOS 
    in $\bar{\Lambda}$, $\bar{M}$ space, and the HPD bound $\tilde{\Lambda}>70$ from 
     \cite{LVC:BNSSourceProp:2019}.
    }
    \label{fig:mismatch}
  \end{center}
\end{figure}

Before discussing the impact of waveform systematics in general, we start with a 
more specific question. When computing a waveform with one waveform model
evaluated at some example parameters, and computing the best match with a different
waveform model, how strongly do the parameters differ?
We focus on the idealized case of fixed chirp mass (using the GW170817 value) and zero spin,
working in a parameter space $\bar{M}$, $\bar{\Lambda}$.
Figure~\ref{fig:mismatch} shows the best matches with regards to a few reference 
parameters, comparing the IMRPhenomD\_NRTidal \cite{Dietrich:2017:1706.02969}
and TaylorF2(6PN) \cite{Vines:2011:084051} waveform models.
We note that the two do not share the same point-particle baseline and that 
the TaylorF2 shows relatively large differences to more recent 
models (see also \cite{LVC:BNSSourceProp:2019}). How the true error 
of the most sophisticated models compares to the differences found for this 
example pair is an important open question, but outside the scope of this work.

As one can see, the differences for reference points with zero deformability 
are comparable to the lower credible bounds for the tidal deformability 
given in \cite{LVC:BNSSourceProp:2019} (with narrow spin prior). 
In contrast, the Bayesian lower bounds obtained in 
\cite{LVC:BNSSourceProp:2019} for different waveform models (including 
the two examples used here) agree quite well. This is not necessarily 
a contradiction. As we discussed before, one would expect similar lower bounds
even when low tidal deformabilities cannot be distinguished at all at the given
sensitivity. 
The comparison of statistical parameter 
estimation studies with different waveform models is an important cross-check,
but it should not be used exclusively to estimate the systematic error due to 
the waveform. On the contrary, if lower bounds agree better than the 
uncertainties obtained from the measures discussed here, it should be taken as 
an indication that the lower bounds are not informative.

The comparison of best matches for different waveform models gives a reasonable 
lower limit for the waveform-related systematic parameter uncertainty. We note, 
however, that the parameter difference at the best match could \textit{a priori} be 0 despite 
a large residual mismatch at the best matching parameter. Further, true waveform 
most likely differs from both models. 

We therefore devised a complementary error
estimate that is not based on the best matching parameters, but on the magnitude
of mismatch.
For this, we consider two waveform models $A$,$B$ and two points $P_1, P_2$ in 
parameter space.
The mismatch between waveform model $A$ evaluated at the two points can be used
as a measure of how mismatch changes with distance in parameter space.
The mismatch between waveform models $A$ and $B$ evaluated at the same point $P_1$ 
can be used to give us a scale for mismatches between waveform models.
Comparing the two, we define a distance in parameter space 
that is symmetric in the two waveform models,
\begin{align}\label{eq:err_dist}
\begin{split}
D(P_1, P_2) &= 
\frac{\min(F(a_1, a_2),
      F(b_1, b_2))}{\max(F(a_1, b_1), 
                               F(a_2, b_2))} 
\end{split},
\end{align}
where $a_i$ and $b_i$ are the waveforms obtained with models $A$, $B$ at points $P_i$.

If two parameter sets differ by less than $D<1$, they cannot be distinguished reliably 
by measurement as long it is not known if $A$ or $B$ is the correct model.
In practice, we have to assume that the true systematic error is bounded
below $D<E$ if $D$ is computed for a pair of actual waveform models, 
where $E$ is a guess of how much larger the true error is compared to the differences 
between available models.

The contours $E=1,2$ are shown in \Fref{fig:mismatch}, again 
for IMRPhenomD\_NRTidal and TaylorF2(6PN).
It is worth noting that for reference points with zero deformability,
the differences to the best matching parameters are much smaller
than the extent of the $D=1$ contours. This might indicate that 
the mismatch changes only slowly for low tidal deformability. The slower
the change in mismatch, the larger the errors of the deformability.
The location of the best match is probably not a good error estimate
in this case. 
This concerns studies like \cite{Favata:2014:101101}, which use
the best match location to estimate systematic errors, and which might 
similarly underestimate the true parameter error.

As a more conservative error estimate, we can use the maximum of the two 
heuristic measures.
From this, we conclude that for mass ratios in the range $0.7\le q \le 1$
and fixed chirp masses around the one for GW170817, 
\emph{we cannot distinguish a deformability $\bar{\Lambda} \approx 200$ from the 
binary black hole case} $\bar{\Lambda}=0$, unless the waveform model employed 
is known to be more accurate than the difference between
the IMRPhenomD\_NRTidal and TaylorF2 models, i.e. $E<1$.

For comparison, \cite{SoumiDe:2018} claims evidence for finite-size effects
based on lower limits $\tilde{\Lambda} > 38,54,45$ obtained
for the three different priors considered therein (those limits are the ones
obtained without applying a manual lower cutoff motivated by causality constraints 
in NS models).
In light of our waveform systematic discussion, this claim implies a great deal of 
trust in the TaylorF2 waveform model used in \cite{SoumiDe:2018}.
Similar arguments apply to \cite{LVC:EOSPaper:2018}, which does not 
provide a limit for $\tilde{\Lambda}$,
but lower limits for the individual radii. Using the same universal compactness relations
as \cite{LVC:EOSPaper:2018}, we find
that $\bar{\Lambda} = 200$ corresponds to a NS radius of $10.4 \,\mathrm{km}$ for 
the equal mass case, which is again larger than the lower limit $R>9.1 \usk\kilo\meter$ 
given in \cite{LVC:EOSPaper:2018}.

\subsection{Improving GW analysis}
\label{sec:improved_analysis}

In the following, we propose an alternative to the universal relation and EOS-agnostic
analysis methods that avoids the shortcomings discussed in Secs.~\ref{sec:ideal}--\ref{sec:real}.

The first ingredient is to prescribe the priors for $\bar{\Lambda}$ and $\bar{S}$ directly.
In contrast to the universal-relation-based parametrization used in \cite{LVC:EOSPaper:2018},
one retains direct control over the $\bar{\Lambda}$ prior, avoiding also the gaps discussed
in \Sref{sec:real}.
It also avoids correlation with the prior for the mass ratio. 
One can simply prescribe a flat prior for $\bar{\Lambda}$. This avoids disfavoring low values 
of $\bar{\Lambda}$ prior, contrary to the EOS-agnostic flat $\Lambda_1,\Lambda_2$ prior.
We stress that this flat prior is still an \textit{ad-hoc} choice not based on any independent observations. 
One should always check how much results depend on such a prior.

When using $\bar{\Lambda}$ and $\bar{S}$ as independent variables, 
the individual deformabilities are given by
\begin{align}
\Lambda_1 &= q^{-\bar{S}/2} \bar{\Lambda}_0, & \Lambda_2 &= q^{\bar{S}/2} \bar{\Lambda}_0, 
\end{align}
where
\begin{align}
\bar{\Lambda}_0 &= \frac{\bar{\Lambda}}{w_1(q) q^{-\bar{S}/2} + w_2(q) q^{\bar{S}/2}}.
\end{align}
The upper cutoff for the $\bar{\Lambda}$ prior can be set to any reasonable cutoff for
the tidal deformability of a single NS. From  Eqs.~(\ref{eq:lambda_eff}) and~(\ref{eq:lambda_bar}) 
it is easy to show that 
$\bar{\Lambda} \le \max(\Lambda_1, \Lambda_2)$ for all mass ratios.

The main advantage of specifying the prior directly for $\bar{\Lambda}$ is that the 
prior for the second parameter (here $\bar{S}$) does not influence the posterior of $\bar{\Lambda}$.
The reason is that $\bar{\Lambda}$ is the only relevant parameter for the GW signal,
unless $\bar{S}$ reaches values that would cause significant corrections in the waveform 
model employed.
Although the corrections may never become significant unless the signal-to-noise ratio
is much better than GW170817, this needs to be validated in any careful analysis. 
We propose performing parameter estimation with a prior in $\bar{S}$ that is much wider 
than any model assumptions one intends to make. 
Most likely, it will turn out that various model assumptions of interest 
constrain $\bar{S}$ much more than the observational data (at least for any detection 
comparable to GW170817). 
This simplifies the parameter
estimation task considerably because model assumptions for $\bar{S}$ do not need to be
included during the costly parameter estimation computations, but can be incorporated in a 
postprocessing step, as we detail later.

The second ingredient is to express EOS constraints inferred from the measurement
in terms of credible regions in the parameter space $M_c, \bar{M}, \bar{\Lambda}$.
Compatibility with a given EOS can be tested by comparing to the two-dimensional 
subspace allowed by the EOS. For a detection similar to GW170817, the chirp mass is
known very accurately and can be treated as fixed. In that case, the EOS surface
reduces to one-dimensional curves in ($\bar{M}, \bar{\Lambda}$) space, as in 
\Fref{fig:lambda_m_vs_M_m_all}.
This approach also allows comparing to the mixed binary hypothesis with given EOS in a unified way, without
resorting to separate parameter estimation studies with parameter spaces of different 
dimensionality, or to the EOS-agnostic method. 
It also allows one to judge the compatibility of arbitrary EOS models
without new parameter estimation computations.

One should avoid providing results from GW data primarily in terms of Bayesian 
credible regions on the NS mass-radius relations. 
Although it simplifies comparison to astronomical observations of isolated pulsars 
and allows one to use known mass-radius curves for EOS models, it also
requires additional model-dependent assumptions, e.g., universal relations.
As we have seen, it is nontrivial to assess how much the final result depends on
measurement and how much on model assumptions. In contrast, $M_c, \bar{M}$, and 
$\bar{\Lambda}$ are directly constrained by the GW signal and do not require model 
assumptions. Presenting results in this parameter space seems the cleanest line of
separation between the overlapping problems of GW data analysis on one side and 
NS structure models on the other.

When results on individual NS properties are presented, the unavoidable direct dependency 
on model-specific constraints of $\bar{S}$ needs to be quantified. Otherwise, it is impossible
to judge how much results depend on assumptions and how much on measurement constraints.
We note that current models only yield approximate ranges for $\bar{S}$, possibly as a function 
$M_c, \bar{M}$, and $\bar{\Lambda}$,
but not a well-motivated shape of the prior distribution. 
As we remarked earlier, $\bar{S}$ is typically much more constrained
by assumption than by measurement. 
It is therefore not meaningful to repeat expensive 
parameter estimation computations with different priors for $\bar{S}$. 
Instead, one can apply cuts to the posterior, limiting $\bar{S}$ to the range allowed
by a given model assumption.

To get an estimate on the impact of model assumptions constraining $\bar{S}$ to a range,
on can further marginalize the restricted posterior over $\bar{S}$, and then derive two new 
posteriors for each of the individual deformabilities $\Lambda_1, \Lambda_2$ by setting 
$\bar{S}$ to the extreme values allowed by the model constraints.
Repeating this for different models provides a computationally inexpensive 
way to separate model uncertainties and measurement error.

What type of credible intervals, e.g., two-sided symmetric, highest posterior density (HPD), 
or upper limits,
are meaningful to provide for the statistical error 
depends on the posterior shape. For the tidal deformability, there likely exists a 
reliable one-sided upper limit, since larger tidal effects are increasingly 
easier to detect. For a detection similar to GW170817, lower limits require more care,
as argued in previous sections. We note that using a HPD interval, as done in 
\cite{LVC:BNSSourceProp:2019}, does not automatically 
yield a meaningful choice. For example, after adding an infinitesimal perturbation
to a flat posterior, the HPD choice can result in anything from one-sided lower to
one-sided upper limits.

In general, one should compare the posterior and prior in detail, also 
distinguishing the impact of mass constraints and deformability constraints.
This is simpler when using a flat prior in $\bar{\Lambda}$ as proposed, 
since the latter is not correlated with the masses. 
It is then not necessary to restrict 
the prior to the measured mass range for the comparison, in contrast to the universal
relation method (see discussion in \Sref{sec:real}).

A simple diagnostic check is given by the uninformative detector test introduced 
in \Sref{sec:real}. The cutoff value 800, motivated by GW170817, should of 
course be replaced with the one-sided upper limit for the posterior at hand.
Lower bounds should be interpreted with care if the lower bound 
is not larger than the one given by the uninformative detector test. In such cases, 
lower bounds should not be regarded as observational constraints, at least not 
without a more careful analysis.
Even when passing this test, one should 
still perform Bayesian model selection as in \cite{LVC:EOSModelSel:2019}, since credible 
intervals do not imply statements on the likelihood of specific 
models (see Sec.~\ref{sec:intro}).

The waveform systematics should not be assessed solely by comparing 
parameter estimation results using different waveforms.
It is possible that statistical bounds agree much better than the systematic 
uncertainty, if the bounds are determined mainly by the prior.
One should use mismatch computations as shown in \Sref{sec:syserr} as a guideline,
and not trust lower limits below values for which the waveform differences
are potentially larger.

We conclude with some remarks on incorporating EM counterparts. Improvements in 
numerical modeling of counterparts likely allows one to obtain robust 
predictions of EM counterparts.
However, there is no reason to assume that the parameter
combinations that are most constrained by EM counterparts
will bear any similarity to those combinations constrained by the GW signal.
In particular, it might not be possible to express the impact of the EOS on the 
kilonova signal solely in terms of tidal deformability. In \cite{Radice:2018:852L}, 
a few numerical simulations were used to propose a loose correlation between effective tidal 
deformability and an upper limit for mass ejection. 
A later work \cite{Kiuchi:2019:876L} considered 
different EOS and found counterexamples, also demonstrating a large impact of
the mass ratio. For reliable results, it is likely necessary to consistently 
use a parameter space with enough parameters to uniquely characterize 
both EM and GW observable signals. 
One way to construct such a space is to use
a parametrized EOS both for GW data analysis and numerical simulations predicting
EM counterparts (the latter also requires one to parametrize the thermal part).
Since numerical simulations are computationally expensive,
analytic fits cannot be avoided completely. In order to prevent conceptual 
problems similar to those we discussed for the universal relations, 
they should however be based on data points in the parameter region relevant for a given event.

\section{Summary}
\label{sec:concl}

In this work, we investigated the question whether results presented in 
\cite{LVC:EOSPaper:2018,SoumiDe:2018,LVC:BNSSourceProp:2019} could be interpreted
as observational evidence for tidal effects during the merger that caused 
gravitational-wave event GW170817. 
Such a claim was made in \cite{SoumiDe:2018}, and also in 
\cite{LVC:BNSSourceProp:2019} assuming low NS spins.
Under the assumption that both objects are NSs, the authors of
Ref.~\cite{LVC:EOSPaper:2018} claim a measurement of lower bounds for the NS radii. 
Since those are derived by feeding the tidal deformabilities into a given monotonic 
analytic expression, this claim implies a measurement of lower limits for tidal effects.

All of those works use Bayesian statistics, but do not satisfactorily address the 
question of how much the results are already determined by the prior assumptions, and 
how much the observational data contribute.
We discussed the prior probability distributions in detail. 
Their features are based largely on arbitrary choices,
which is particularly problematic for lower limits on tidal effects, 
since small tidal effects have a weaker impact on the measured signal.

Since the detector cannot distinguish small tidal effects as well as large ones,
we discussed the idealized case of a detector that cannot distinguish
tidal effects at all if they are below published upper credible bounds.
Using such manifestly uninformative hypothetical measurement with the 
same prior assumptions, we still obtained lower limits comparable 
to the published ones.
Our findings indicate that the Bayesian lower credible bounds in
\cite{LVC:EOSPaper:2018,SoumiDe:2018,LVC:BNSSourceProp:2019}
might largely reflect prior assumptions.
They should not be regarded as observational evidence, at least not without 
a more detailed analysis.

During our study, we also found several technical and conceptual problems. 
The two studies 
\cite{LVC:EOSPaper:2018,SoumiDe:2018} are using model assumptions based on 
universal relations in a way that is not self-consistent. In short, those 
relations already imply much larger lower limits just based on the measured
mass range, and hence cannot be used to measure finite tidal effects.
While the use of universal relations in \cite{SoumiDe:2018} has no consistent
motivation when applied to low tidal deformabilties for the mass range of GW170817, 
it still allows such values.
The universal relations used in \cite{LVC:EOSPaper:2018} on the other hand 
completely exclude values below a limit depending on mass ratio. This limit
is not a physical assumption, but results from extrapolating a complicated 
fit function given by a fraction of polynominals.

Regarding the alternative approach of parametrized EOS discussed in 
\cite{LVC:EOSPaper:2018}, we presented arguments suggesting that the impact
of the prior is not sufficiently studied to claim observational evidence.
However, our discussion is not conclusive since we did not quantify the impact.

Our reassessment of the EOS-agnostic study \cite{LVC:BNSSourceProp:2019} also 
revealed some problems. On the conceptual level, we argued that the use of a 
rescaled posterior, which mimics a flat prior in the effective tidal 
deformability, in effect corresponds to a prior that not based on well-defined 
physical assumptions.
Moreover, it is ambiguous how the prior in the multidimensional parameter
space should be rescaled to achieve a given marginalized prior. 
On the technical level, we investigated 
the correlation with the mass ratio and found that the full rescaled prior
from \cite{LVC:BNSSourceProp:2019} is not flat at all when restricted to the 
measured mass range, defeating the apparent purpose of the procedure.

For use in future studies, we sketched a GW data analysis strategy that sidesteps
some of the emerging problems. The main source of complications seems to be
the inclusion of quantities that cannot be measured, necessitating the use of 
model assumptions, and complicating the interpretation of results considerably.
The main idea of our proposal is to restrict the data analysis to quantities that 
actually affect the waveform and can hence be measured. Compatibility with model
assumptions can be tested in a second step, as well as derivation of quantities 
that require further model assumptions. 
Our strategy also simplifies quantifying
the impact of other model assumptions independently.

We demonstrated explicitly how to test compatibility with EOS models.
This also showed that, based on a single event with masses both compatible with 
NS, it is fundamentally impossible to distinguish a BNS merger 
with one EOS from a BH-NS merger with a different EOS, as long as tidal effects enter
the waveform model only via the effective deformability.
We also provided examples demonstrating that the maximum mass of NSs
cannot be inferred from the inspiral GW signal alone.

Another aspect particularly important for measuring finite tidal effects
is the impact of systematic waveform model uncertainty.
This was addressed in \cite{LVC:BNSSourceProp:2019} by repeating the full 
statistical analysis with different waveform models. 
A similar approach
is to perform parameter estimation studies based on 
one waveform model on artificial data constructed using a different waveform
model. We argued that this 
is only a valid error estimate for credible bounds if they depend on the 
observational data, instead of the prior. We derived an independent estimate
by directly computing differences between waveforms, using the mismatch 
measure relevant for GW data analysis for LIGO-type detectors.
The results from this estimate indicate that also previous error estimates
based on best-matching waveforms, e.g., \cite{Favata:2014:101101}, might be too optimistic.
We find that the uncertainties expected from differences of current waveform
models are larger than the lower limits inferred for GW170817. 

Conclusive 
observational evidence of finite tidal effect therefore requires  
better knowledge of the waveform accuracy, as well as another merger event
with either a larger signal-to-noise ratio or involving objects with larger 
deformabilties. For standard NS models, that means
less massive NS. However, the mass range of observed NS is rather limited
so it is not guaranteed that such an event will occur.

\acknowledgements
We thank Soumi De, Duncan A.\ Brown, and Katerina Chatziioannou for useful 
discussions, and Philippe Landry for a careful review of the initial draft of 
this paper. This work was supported by the Max Planck Society's Independent 
Research Group Programme. 
In this work, we use a neutron-star model database computed on
the Holodeck cluster at the Albert Einstein Insitute in Hannover. 

\bibliographystyle{apsrev4-1-noeprint}
\bibliography{article}

\end{document}